\documentclass[12pt]{article}
\usepackage{epsfig}
\usepackage{cite}
\newcommand{\be}{\begin{equation}}
\newcommand{\ee}{\end{equation}}

\begin{document}

\title{Phenomenological analysis of  experimental data on
$\eta$-photoproduction on protons\footnote{The contributed paper to MENU2001, 
Washington, USA, 26-31 July}}
\author{E.M.Leikin, E.V.Balandina, N.P.Yudin\\
\and Nuclear Physics  Institute, Moscow State University, Moscow,
Russia}
\maketitle

\begin{abstract}
The results of linear regresion of differential cross sections, $\Sigma$- and
$T$-asymmetries of $\eta$-photoproduction on protons in energy region from
threshold up to 1 GeV are presented. Serious contradictions between angular
distributions measured in different laboratories are revealed. The energy dependance
of regression coefficients may be due to transition from energy region of
$S_{11}(1535)$ and $D_{13}(1520)$ to  energy region of $D_{15}(1675)$ and
$F_{15}(1680)$ resonances.\\
\end{abstract}

During the past years  $\eta$-photoproduction on protons has
attracted increasingly high interest. This is due not only because this item is a new
physical phenomenon different from photoproduction of pions but
mainly because $\eta$-photoproduction should proceed through the small number of
nucleon resonances. Even in energy region up to 1 GeV there will
be not too many overlapping resonances that permits to extract 
 reliable information on resonance parameters from experimental
data.

Complete phenomenological analysis of experimental data on photoproduction, as
a rule, encounters a number of problems, e.g. solving  of nonlinear
equations, removal of continuous and discrete theoretical ambiguities,
elimination of experimental ambiguities, etc. 
Analysis of experimental data may be naturally divided in two stages
\cite{1}. The first is the linear regression which provides the
information about the number of partial waves that contribute to the measured
experimental characteristics of process and provides  information on the
resonances concerned. The linearity of the model used ensures that 
 the 
estimates of regression coefficients are unbiased. The second is to  determine the
multipole amplitudes.

\begin{figure}[h]
\vspace*{-2cm}
\epsfig{file= 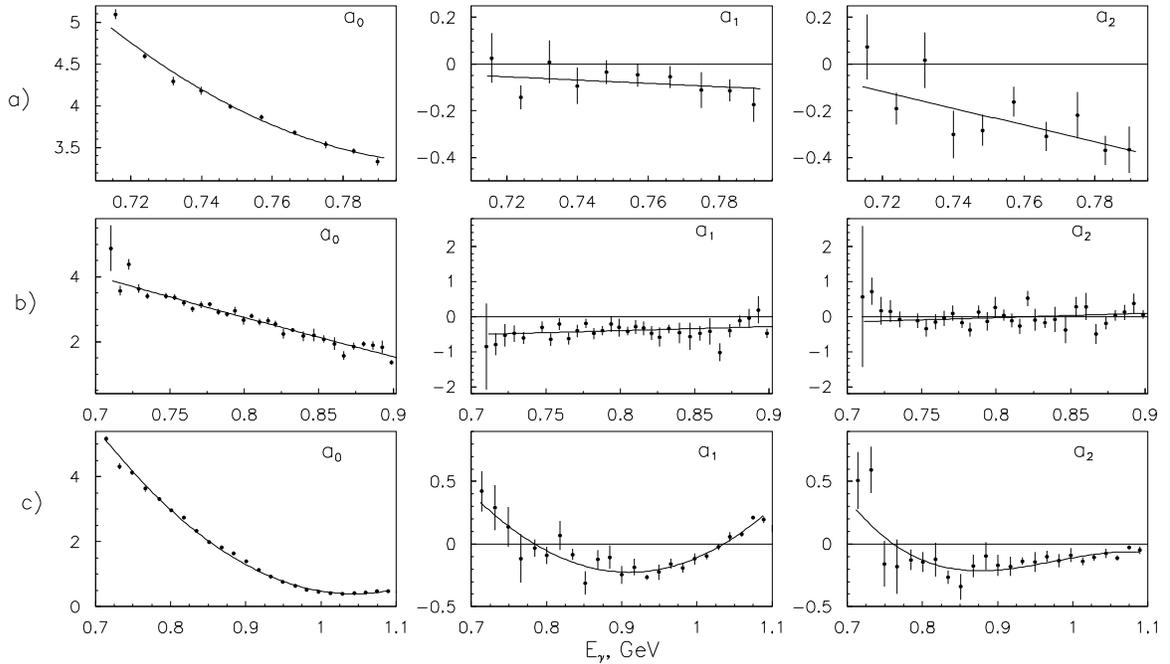}
%\hfill
\vspace*{-2cm}
\caption{\label{fig1}
Coefficients $a_0,\ a_1,$ $ a_2$ in expansion  (1a). Data are taken (a) from 
  \protect\cite{2}, (b) from \protect\cite{3}, (c) from \protect\cite{4}; lines show the results of 
the fit.}
\end{figure}

\begin{figure}[h]
\vspace*{-2cm}
\epsfig{file= 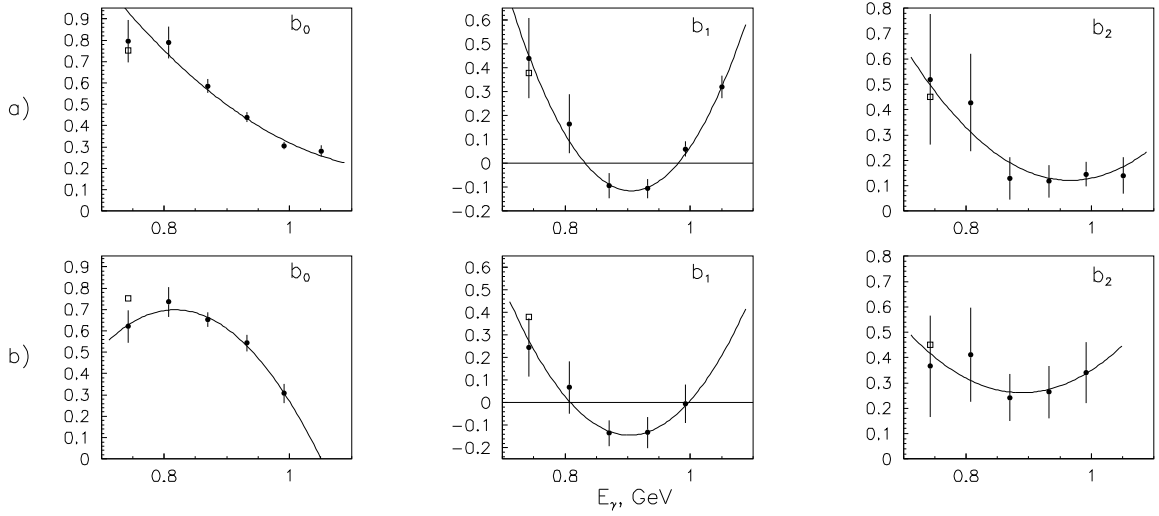}
\vspace*{-2cm}
\caption{\label{fig2}
Coefficients $b_0,\ b_1,$ $ b_2$ in expansion  (1b). Data for $\Sigma$ are 
taken from \protect\cite{5}.
(a) $d\sigma/d\Omega$  from \protect\cite{4}, (b) $d\sigma/d\Omega$  
from \protect\cite{3}. Square symbols: $d\sigma/d\Omega$ from \protect\cite{2}.}
\end{figure}

\begin{figure}[h]
\vspace*{-2cm}
\epsfig{file= 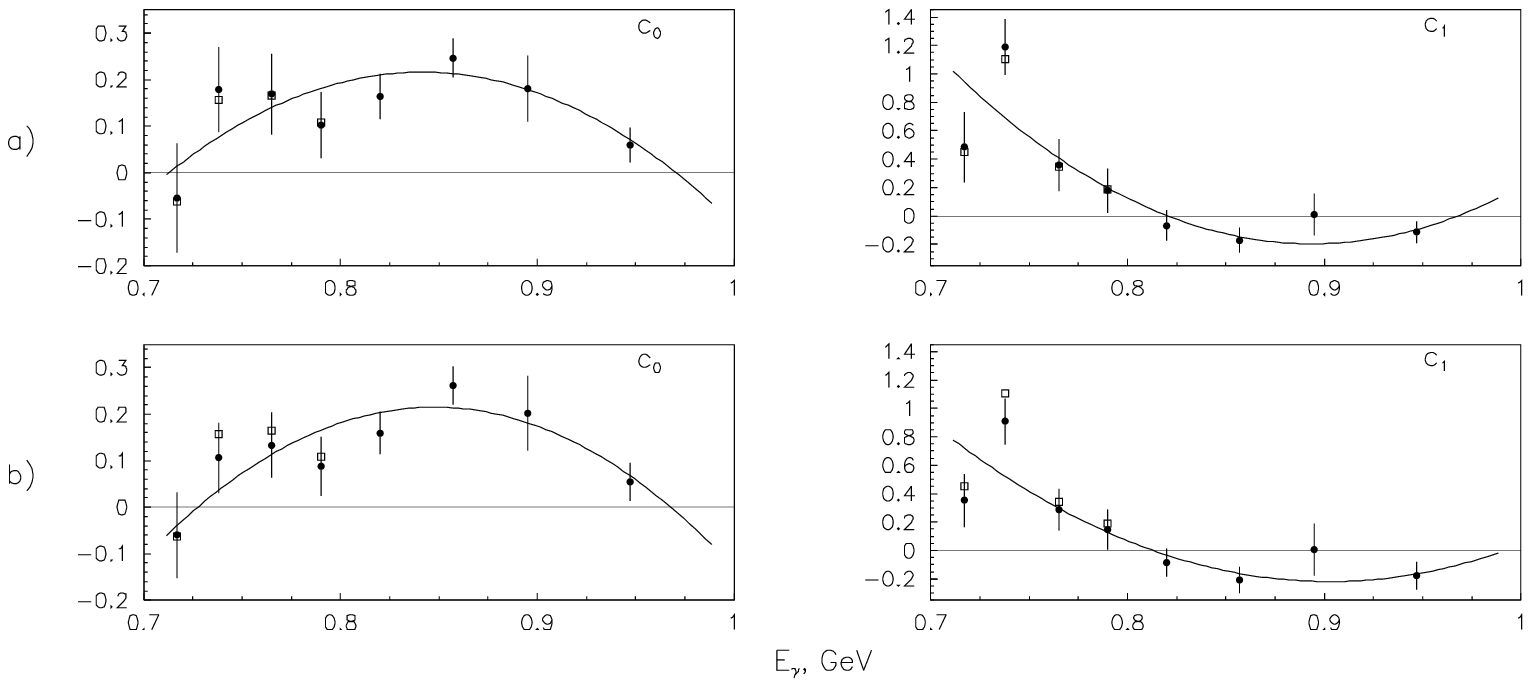}
%\hspace*{-2cm}
%\hfill
\vspace*{-2cm}
\caption{\label{fig3}
Coefficients $c_0,\ c_1$ in expansion  (1c). Data for $T$ are taken 
 from \protect\cite{6}.
(a) $d\sigma/d\Omega$  from \protect\cite{4}, b) $d\sigma/d\Omega$  
from \protect\cite{3}. Square symbols: $d\sigma/d\Omega$ from \protect\cite{2}.}
\end{figure}

This paper is confined to the first stage of analysis. We have analysed all known experimental data on differential
cross sections (angular distributions) of process $\gamma p \rightarrow \eta p$
\cite{2,3,4}, and also the data on polarization observables, i.e.
angular distributions of asymmetry $\Sigma,$ measured with linear polarized beam
\cite{5} and angular distributions of asymmetry $T,$ measured on a polarized
target \cite{6}. The energy independent analysis consist in expanding angular
distribution of the observables at definite energy using Legendre polinomials. To
find how many terms in this expansion provide the best description of data
standart statistical procedures including the Fisher criterion were used. 
Unlike the energy dependent analysis that is based on parametric
models  and, generally, gives biased estimates, energy independent analysis relies on  nonparametric model that provides 
unbiased estimates.
Expansion of the observables and corresponding statistics are:
$$
{ k\over q } \, {{d\sigma(\theta)}\over{d\Omega}} =\sum a_n\,P_n(\cos \theta)\,,
\hfill \eqno\hbox{(1a)}
$$
$$
{ k\over q } \, {{d\sigma}\over{d\Omega}}\,{1\over{\sin^2\theta}}\,\Sigma
=\sum b_n\,P_n(\cos \theta)\,,
\hfill \eqno\hbox{(1b)}
$$
$$
{ k\over q } \, {{d\sigma}\over{d\Omega}}\,{1\over{\sin \theta}}\,T
 =\sum c_n\,P_n(\cos \theta)\,.
\hfill \eqno\hbox{(1c)}
$$
Multipole decomposition of coefficients $a_n,\ b_n,\ c_n$ up to terms
$E_{3-}$ and $M_{3-}$ may be found in \cite{7}.
In all cases the best
description of experimental data on $d\sigma/d\Omega,$ 
were obtained with three terms 
of the expansion. 
The dominance of $s$-wave, the
coefficient $a_0,$ was already pointed out \cite{2,8}. However,
 the coefficients $a_1$ and $a_2$ connected,
correspondingly, with the $sp$- and $sd$-interferences demonstrate the existence
of serious contradictions between  results in \cite{2,3,4}. This is
 also diplayed by
Fig.~\ref{fig1}.
%\vspace*{-2cm}
Since the observables $d\sigma/d\Omega,$ $\Sigma,$ $T$ were
measured  at different energies and angles to form the
statistics with $\Sigma$ and $T$ we  used interpolated values of $d\sigma/d\Omega.$
The polarization statistics $\Sigma$ and $T$ were analysed with both
$d\sigma/d\Omega$ obtained in the same laboratory and  $d\sigma/d\Omega$
from another laboratories.  To get the description of $\Sigma (\theta)$ it was
necessary to keep  three terms in expantion. For the $T(\theta)$ it was sufficient to
keep two terms. The energy dependence of $b_n$ and $c_n$ is shown on
Fig.~\ref{fig2} and \ref{fig3}.

\noindent Contradictions between  angular distributions
obtaned in  different laboratories are not  reflected in the general behavior
of the  coefficients $b_n$ and $c_n.$ In other words, these
contradictions between $d\sigma/d\Omega$ do not appear in polarizational
observables.

It seems to be instructive to consider the energy behaviour of coefficients
$a_1,\ b_1,\ b_2$ and $c_0,\ c_1.$ 
The change of energy dependence of this coefficients at 0.9 GeV might
indicate the change of regime of the process. For instance, the decrease of
$a_1$ \cite{4} from the threshold to 0.9 GeV may be due to the damping of $s$-wave and to
weakening of $sp$-interference. The further rise of $a_1$ may be related
with the contribution of higher partial waves. The decrease of $b_2$ at energies
below 0.9 GeV may be related with resonance $D_{13}(1520)$; the growth of
$b_2$ at energies 0.9--1.1 GeV may be due to influence of resonances
$D_{15}(1675)$ and $F_{15}(1680).$ The interference of $d$- and $f$-waves should
lead to the shift of angular distribution $\Sigma (\theta)$ to the smaller
angles in the CM system as really seen in experiment \cite{5}. The
behaviour of $b_1$ at energies higher than 0.9 GeV can be attributed to
$sf$-interference and so on.

Thus,   energy dependance of regression coefficients found in our analysis may
be due to the transition from  energy region of $S_{11}(1535)$ and
$D_{13}(1520)$ to  energy region of $D_{15}(1675)$ and $F_{15}(1680).$

\end{document}